\documentstyle[epsf,eqsecnum,floats,preprint,aps]{revtex}

\tighten

\begin{document}

\newcommand{\Bd}{{\dot B}}
\newcommand{\Cd}{{\dot C}}
\newcommand{\fd}{{\dot f}}
\newcommand{\hd}{{\dot h}}
\newcommand{\ep}{\epsilon}
\newcommand{\vp}{\varphi}
\newcommand{\al}{\alpha}
\newcommand{\be}{\begin{equation}}
\newcommand{\ee}{\end{equation}}
\newcommand{\bea}{\begin{eqnarray}}
\newcommand{\eea}{\end{eqnarray}}
\newcommand{\PSbox}[3]{\mbox{\rule{0in}{#3}\includegraphics{#1}\hspace{#2}}}

\title{Topological Inflation with Multiple Winding}

\author{Andrew A. de Laix\footnote[1]{\tt aad4@po.cwru.edu}, 
Mark Trodden\footnote[2]{\tt trodden@theory1.phys.cwru.edu} and 
Tanmay Vachaspati\footnote[3]{\tt txv7@po.cwru.edu.}}

\address{~\\Particle Astrophysics Theory Group \\
Department of Physics \\
Case Western Reserve University \\
10900 Euclid Avenue \\
Cleveland, OH 44106-7079, USA.}

\maketitle

\begin{abstract}

We analyze the core dynamics of critically coupled, superheavy gauge 
vortices in the (2+1)
dimensional Einstein-Abelian Higgs system. By numerically solving the 
Einstein and field equations for various values of the symmetry breaking
scale, we identify the regime in which static solutions cease to exist
and topological inflation begins. We explicitly include the topological
winding of the vortices into the calculation and extract the dependence
on the winding of the critical scale separating the static and inflating 
regimes. Extrapolation of our results  suggests 
that topological inflation might occur within high winding strings formed
at the Grand Unified scale.

\end{abstract}

\setcounter{page}{0}
\thispagestyle{empty}

\vfill

\noindent CWRU-P23-97 \hfill Submitted to {\it Physical Review} {\bf D}

\hfill Typeset in REV\TeX

\eject

\vfill

\eject

\section{Introduction}
Two of the most exciting possibilities suggested by modern particle cosmology
are the production of topological defects and the existence of an inflationary
stage in the early universe. In the last few years it has been pointed out 
independently by Vilenkin \cite{AV 94} and by Linde \cite{AL 94}, that these 
two phenomena may be related.
A necessary condition for inflation to occur is that the energy density of the
universe be dominated by the vacuum energy of a homogeneous scalar field, 
called the ``inflaton''. Correspondingly, a central feature of topological 
defects is that their cores are regions of spacetime in which a scalar field 
is forced to sit out of its vacuum manifold, taking a value corresponding to a 
false vacuum. Hence, the cores of topological defects are regions in which
energy density is trapped in the form of the vacuum energy of a scalar field.
Vilenkin and Linde realized that if the energy density trapped in a defect 
core and the core radius itself were large enough, this energy could
be considered uniform (ie. horizon sized), and the core would satisfy the 
conditions for inflation. This scenario is what is known as
{\it topological inflation}.  The advantage of this implementation is 
that, while traditionally the necessary conditions require the fine tuning of a
scalar potential, here inflation is inevitable if the vacuum manifold of
the theory at high temperatures satisfies a topological constraint. Thus,
the question of initial conditions becomes one of topology.

If defects are produced when a scalar field gets a vacuum expectation
value equal to $\eta$ in the early universe, then an
approximate criterion proposed by Vilenkin and Linde for topological
inflation to occur is $\eta > m_p$, where $m_p$ is the Planck mass. 
(Note that the energy density in the scalar field is proportional to
$\lambda \eta^4$ which is assumed to be less than the Planck energy
density.) The supposition that, under this condition,
time-dependent defect solutions are inevitable, is supported by the 
traditional solutions for the spacetime around, for example, a 
cosmic string. At symmetry breaking scales significantly below the Planck
scale, the spacetime around a static cosmic string is conical \cite{Linet}, 
with a deficit angle proportional to $\eta^2$. However, as $\eta$ increases 
to of order the Planck mass, the deficit angle approaches $2\pi$ and static 
solutions cease to exist \cite{Ortiz}. Recently, the criterion for 
topological inflation has 
been made more precise by numerical simulations of the spacetime structure 
around various defects; gauge monopoles \cite{Sakai}, domain walls 
\cite{Sakaietal}, and global monopoles \cite{{Sakaietal},{C&V 97}}. 
These analyses all focus on defects with unit topological charge, and find 
that in that case, the criterion for topological inflation is 

\be
\eta >\eta_{cr} \equiv 0.33 m_p \ .
\label{etacritical}
\ee
The aim of this paper is first to verify that the above criterion holds
for the important case of gauge cosmic strings, and second, to investigate 
how the criteria for topological inflation depend on the topological charge of 
the defects considered. 

Our motivations are twofold. First, for defects of unit 
winding,~(\ref{etacritical}) implies that, to realize topological inflation, 
we must work very close to the Planck scale, at which our field theories may
not be valid. Further, it is clear that topological 
inflation cannot occur
at the GUT scale for unit winding defects. We hope that, when higher
topological charges are included, these constraints will be alleviated
and the value of $\eta_{cr}$ will decrease. In fact, one might expect such 
behavior from considering the static string solutions, since the deficit 
angle we 
mentioned above is also dependent on the winding $n$, and so static solutions
should cease to exist for lower values of $\eta$ if $n>1$. The spacetime
structure for such higher winding strings is the focus of this paper.

Our second motivation comes from models in which particles are described as
solitons \cite{solitons}, and in particular from the 
{\it dual standard model} \cite{dual}. In this theory, all the standard model
particles arise as monopoles of some bosonic field theory. In such a
model it is natural to wonder what happens to matter at high densities when
the core structure of the solitons becomes important. For example, stars can
be seen as collections of huge numbers ($\sim 10^{57}$) of monopoles. 
If the monopoles become squeezed together tightly enough 
for a large region of the star to be in the false vacuum with high winding, 
might topological inflation occur?

In the present work, fueled by the above considerations, we consider the
simple example of an Abelian-Higgs vortex with winding $n$ in $(2+1)$ 
spacetime dimensions. We do so because it is easier to work with multiple 
winding vortices than the analogous monopoles, although the problem of 
monopoles for 
$n>1$ is under consideration. In the next section we present the model and
the equations of motion we solve. We give our initial conditions and describe
how we expect solutions to behave in some asymptotic regimes. In section III, 
we briefly discuss the implementation of the numerical algorithms we use to 
solve the equations and in section IV we present our results. Section V 
contains a discussion of the results and their implications for topological 
inflation.

\section{The Model}
Consider the Abelian Higgs model with a complex scalar field $\Phi$ and
a $U(1)$ gauge field $A_{\mu}$, coupled to gravity in $(2+1)$ spacetime
dimensions. The action is

\be
S=\int d^3 x\sqrt{-g}\left(\frac{1}{16\pi G}{\cal R} + {\cal L}\right) \ ,
\label{action}
\ee
where ${\cal R}$ is the Ricci scalar and the Lagrangian density for the 
matter fields is

\be
{\cal L} = 
(D_{\mu}\Phi)^*D^{\mu}\Phi -\frac{1}{4}F_{\mu\nu}F^{\mu\nu} -V(\Phi) \ .
\label{lagrangian}
\ee
Here, the covariant derivative is 
$D_{\mu}\Phi=(\nabla_{\mu}+i\ep A_{\mu})\Phi$,
the gauge field strength is 
$F_{\mu\nu}=\partial_{\mu}A_{\nu}-\partial_{\nu}A_{\mu}$, and the scalar
potential is

\be
V(\Phi) = \frac{\lambda}{4}(\Phi^*\Phi - \eta^2) \ ,
\label{potential}
\ee
with $\lambda$, $\ep$ constants and $\eta$ a mass scale.

In cylindrical polar coordinates ($r$, $\theta$), we make the usual 
Nielsen-Olesen string ansatz for the fields

\bea
\Phi(x) & = & f(r,t)e^{in\theta} \ , \nonumber \\ 
A_{\theta} & = & \frac{1}{\ep}[h(r,t)-n] \ , \nonumber 
\eea
where the integer $n$ is the winding number of the string. 
The metric ansatz is

\be
ds^2 = dt^2 - e^{B(r,t)}dr^2 - e^{C(r,t)}r^2 d\theta^2 \ ,
\label{metric}
\ee
where $B$ and $C$ are functions of the radial coordinate and time.

This ansatze leads to four Einstein equations and two field equations for
a total of four unknown functions: $B$, $C$, $f$ and $h$. Two of the six
equations are first order in time derivatives and are the constraint
equations.
The four equations we solve are:

\bea
e^B(-2{\ddot C} & - & \Cd^2) = \nonumber \\
&  & \frac{32\pi}{m_p^2}\left[
e^B\fd^2 + f'^2 +\frac{e^{B-C}}{2\ep^2 r^2}\hd^2 +
\frac{e^{-C}}{2\ep^2 r^2}h'^2 -\frac{e^{B-C}}{r^2}h^2f^2 -
\frac{\lambda}{4}e^B(f^2-\eta^2)^2\right] \ ,
\label{rreqn}
\eea

\bea
e^C(-2{\ddot B} & - & \Bd^2) = \nonumber \\
&  & \frac{32\pi}{m_p^2}\left[
e^C\fd^2 - e^{C-B}f'^2 -\frac{1}{2\ep^2 r^2}\hd^2 +
\frac{e^{-B}}{2\ep^2 r^2}h'^2 +\frac{1}{r^2}h^2f^2 -
\frac{\lambda}{4}e^C(f^2-\eta^2)^2\right] \ ,
\label{ththeqn}
\eea

\bea
{\ddot f}-e^{-B}f'' +\frac{e^{-C}}{r^2}h^2f +\frac{1}{2}(\Bd+\Cd)\fd -
e^{-B}\left[\frac{1}{r} +\frac{1}{2}(B'+C')\right]f' +
\frac{\lambda}{2}(f^2-\eta^2)f = 0 \ ,
\label{feqn}
\eea

\bea
{\ddot h}-e^{-B}h'' -\frac{1}{2}(\Cd-\Bd)\hd +\frac{e^{-B}}{2}(C'+B')h' +
e^{-B}\frac{h'}{r} +2\ep^2 f^2h = 0 \ ,
\label{heqn}
\eea
where a dot (prime) denotes a derivative with respect to time ($r$). At all 
times, we insist that the two additional constraint equations 

\bea
\Bd\Cd + & e^{-B} & \left(-2C'' -C'^2 +B'C' -4\frac{C'}{r} +
2\frac{B'}{r}\right) = \nonumber \\
&  & \frac{32\pi}{m_p^2}\left[
\fd^2 + e^{-B}f'^2 +\frac{e^{-C}}{2\ep^2 r^2}\hd^2 +
\frac{e^{-B-C}}{2\ep^2 r^2}h'^2 + \frac{e^{-C}}{r^2}h^2f^2 +
\frac{\lambda}{4}(f^2-\eta^2)^2\right] \ ,
\label{ttconstraint}
\eea
and

\be
-2\Cd'-\Cd C' +\Bd C' -2\frac{\Cd}{r}+2\frac{\Bd}{r} =
\frac{32\pi}{m_p^2}\left[2\fd f' +\frac{e^{-C}}{\ep^2r^2}\hd h'\right]
\label{trconstraint}
\ee
are satisfied.

Now consider the initial conditions for these equations. We begin with
a cylindrically symmetric string configuration for the fields $f$ and $h$, 
which is initially static. We define the metric to be initially flat but with
non-vanishing first time derivatives. For the metric, these conditions are
simply implemented as

\be
B(r,0)=C(r,0)=0 \ ,
\label{BCics}
\ee
with $\Bd(r,0)$ and $\Cd(r,0)$ obtained from the constraint 
equation~(\ref{ttconstraint}). For the fields, however, we need the initial
profile functions $f(r,0)$ and $h(r,0)$. To simplify this process, we work 
in the {\it Bogomolnyi limit} defined by

\be
\beta \equiv \frac{\lambda}{2\ep^2} = 1 \ .
\label{bogomolnyi}
\ee
In this case, there are no forces between static strings and the energy
saturates a topological bound. Since we are primarily concerned with
gravitational effects, we do not expect results for $\beta \neq 1$ to be
significantly different. In the Bogomolnyi limit, the static field equations 
reduce to the two first order equations

\be
f'+e^{(B-C)/2} \frac{hf}{r} = 0 \ ,
\label{bogf}
\ee

\be
h'+\frac{\lambda}{2}r e^{(B+C)/2}(f^2-\eta^2) = 0 \ ,
\label{bogh}
\ee
which can be solved numerically.

Our procedure is as follows. We first complete our initial conditions by
solving~(\ref{bogf},\ref{bogh}) subject to the boundary conditions

\bea
\lim_{r\rightarrow \infty} f(r)& = & \eta \nonumber \\
h(0) & = & -n \ .
\label{bogics}
\eea
We then solve 
equations~(\ref{rreqn},\ref{ththeqn},\ref{feqn},\ref{heqn}) with the 
initial conditions we have just described. Throughout the evolution we verify 
that the constraint equations~(\ref{ttconstraint},\ref{trconstraint}) are 
satisfied at each step as a check of our numerical scheme. 

For a given topological charge $n$, this procedure is performed over a range 
of values of the symmetry breaking scale $\eta$. We define a solution 
exhibiting topological inflation to be one for which the total physical
volume, $V_*(t)$, in the core of the defect is increasing exponentially. We 
define $V_*(t)$ by

\be
V_*(t) \equiv 2\pi\int_0^{r_*(t)} dr\, r\exp\left[\frac{B(r,t)+C(r,t)}{2}
\right] \ ,
\ee
where the core radius, $r_*(t)$ is defined by

\be
f[r_*(t)] \equiv \frac{\eta}{2} \ .
\ee
Determining the
functional form of $\eta_{cr}(n)$ is the central result of this paper.

It is a useful check of our results that one may simply derive an upper
bound for the expansion rate in the core. Assume that inflation
occurs and that the metric components $B$ and $C$ become very
large compared to other fields in the core of the defect. Further,
assume that only the vacuum energy of the scalar field is important
in the core, {\it i.e.} that $\phi=0$ with no derivatives important there.
In this approximation, the equations of motion for the metric simplify
dramatically and are easily solved to give

\be
B(0,t) \sim C(0,t) \sim (8\pi\lambda)^{1/2} 
\left(\frac{\eta}{m_p}\right)^2 t \ .
\ee
Thus, any inflationary behavior we observe should have an associated
Hubble constant $H$ that satisfies

\be
H < (8\pi\lambda)^{1/2} 
\left(\frac{\eta}{m_p}\right)^2
\ee

Our intuition for believing that higher topological charges will alleviate the
high symmetry breaking scales required for topological inflation
comes from two sources. First,
consider the asymptotic form of the metric for static cosmic string
solutions (in 2+1) dimensions

\be
ds_{2+1}^2 = dt^2 - dr^2 - r^2d{\tilde\theta}^2 \ ,
\label{staticmetric}
\ee
where ${\tilde\theta}$ is the angle in a locally Minkowski but globally
conical spatial section, taking values in the range

\be
0 \leq {\tilde\theta} < 2\pi\left(1-4|n|\frac{\eta^2}{m_p^2}\right) \ .
\ee
For strings of unit winding, this metric is applicable as long as the
deficit angle is less than $2\pi$, that is, for symmetry breaking 
scales $\eta \ll m_p$. However, for higher winding, static solutions 
cease to exist for 

\be
4|n|\left(\frac{\eta}{m_p}\right)^2 < 1 \ .
\ee
Thus, we expect that asymptotically static solutions become
impossible at a lower critical symmetry breaking scale for defects
with multiple winding and we might guess that the critical value of
$\eta$ at which static solutions cease to exist falls off as
$1/\sqrt{n}$. It is natural to wonder if the same is true in
the core of these defects, although, of course, the absence of static
solutions does not guarantee that the core will inflate.

Second, a perturbative analysis of the matter fields around the center
of such defects demonstrates that defects with $|n|>1$ have a wider
core and higher energy density than the corresponding unit charge
configurations. Both these effects suggest that topological inflation might 
be more easily achieved in high winding defects. 
Unfortunately, it does not seem possible to quantitatively understand
the effect of multiple winding on topological inflation with an analytic 
approach. Thus, here we have solved the system numerically, in the spirit of 
other authors in the unit winding case\cite{{C&V 97},{Sakai},{Sakaietal}}.

\section{Numerical Implementation and Results}
In this section, we present the results of our numerical simulations
of the $(2+1)$ dimensional Einstein-Abelian-Higgs system. The system of
non--linear partial differential equations we study are non-trivial to 
solve numerically. Therefore, before we present our results, let us briefly 
discuss the numerical techniques we use, in the hope that this discussion 
will help others investigating similar problems.  

\subsection{Numerical Implementation}
There are two stages to solving the equations.  The first is to generate the 
field
configurations that will serve as the initial values for the time
dependent evolution equations, and the second is to integrate the
partial differential equations that describe the time
dependence of the fields.  To attack the former, we observe that the
Bogomolnyi equations~(\ref{bogf}) and~(\ref{bogh}) have 
asymptotic solution 

\be
\lim_{r\rightarrow 0} f(r) \sim f_0 r^n \ ,
\ee
where $f_0$ is a constant of integration. This
constant can be used as a free parameter in a shooting method
solution to the boundary value problem.  We find that shooting is
effective in generating accurate solutions out to a radius of 10$\eta$,
sufficient for both fields to reach their $r\rightarrow \infty$ asymptotic
values for the windings we consider.  

With the initial conditions in hand, we must now consider how to
integrate the time dependent partial differential equations.
Typically one replaces derivatives with finite difference approximations.  
For a generic variable $y(r,t)$, one solves for values on a lattice
$y(r,t) \rightarrow y_i^j$, where subscripts indicate the position in
the space lattice and superscripts indicate the location in the time
lattice. Derivatives are replaced with finite difference
approximations, {\it e.g.}
\begin{eqnarray*}
\ddot{y} &\approx& {y^{j+1}_i - 2 y^{j}_i + y^{j-1}_i \over {dt}^2} \\
y'' &\approx& {y_{i+1}^j - 2 y^{j}_i + y^{j}_{i-1} \over {dr}^2} \\
& \vdots &
\end{eqnarray*}
which are second order approximations in $dt$ and $dr$. However,
there is in general no guarantee that a particular
differencing scheme is stable. That is to say, for poor schemes the result
from integrating the difference equations may diverge exponentially
from the true solution.  To test the differencing methods, we may use the
stability analysis for linear equations which is covered in any good
reference on partial differential equations (see {\it e.g}.\ \cite{NR}
chapter 19, and references therein). The results of such an analysis
also provide good
intuition when dealing with the non--linear equations for the metric
and fields we are considering.  For our equations, we find that when
solving for $f$ and $h$, stability is assured if the spatial
derivatives of these fields are evaluated implicitly.  That is, for
the $j^{{\rm th}}$ time step we evaluate the second spatial
derivative of $f$ as 

\be
{f}'' = \left(y_{i+1}^{j+1} - 2 y^{j+1}_i +
y^{j+1}_{i-1}
\right)/ {dr}^2 \ ,
\ee
and similarly for the others.  At each time step, the
implicit scheme gives us a set of 
coupled linear equations for the fields $f^{j+1}_i$ and
$h^{j+1}_i$, solving which reduces to inverting a tri--diagonal matrix, a
standard problem in linear algebra.  We also find that care is
needed when evaluating the metric equations~(\ref{rreqn}) and~
(\ref{ththeqn}).  Both can be written in the form
\begin{equation}
\label{metric_eq_}
	\ddot{C} + \dot{C}^2 = F,
\end{equation}
where $F$ just represents the terms on the right hand side of the
equation.  Treating $\dot{C}$ as an independent variable such that
$\dot{C}^{j+1}_i = \dot{C}^{j-1}_i +2 dt \dot{C}^j_i$, we found that
it is necessary to evaluate $\dot{C}$ implicitly for eq.\
(\ref{metric_eq_}) to be stable, {\it i.e.}
\begin{equation}
\label{metric_eq_diff}
	{\dot{C}^{j+1}_i - \dot{C}^{j-1}_i  \over dt}   + \left(
	\dot{C}^{j+1}_i \right)^2 = F^j_i.
\end{equation}
The above quadratic has two solutions, but one may easily obtain the
right one by taking the limit as $dt \rightarrow 0$ and noting that, for
the correct root, $\dot{C}^{j+1}_i - \dot{C}^{j-1}_i$ should vanish.
These suggestions worked well for us, although they do not represent
the only stable differencing schemes and they may not generalize to
other similar problems.

\subsection{Results}
\begin{figure}
\centerline{
\epsfbox{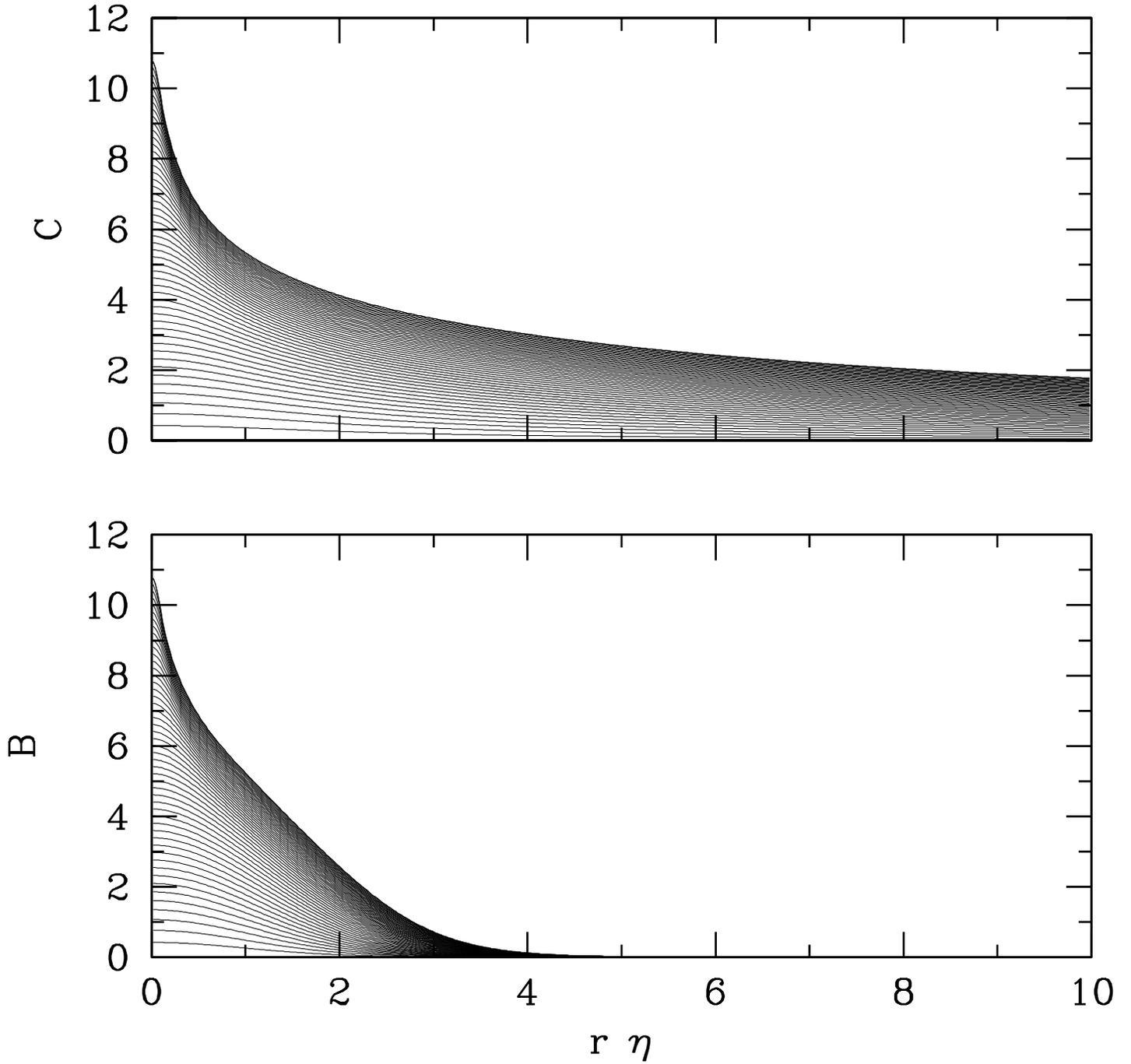}
}
\vspace{0.1in}
\caption{The metric fields $B(r)$ and $C(r)$ for an $n=1$ string with 
symmetry breaking scale $\eta=0.2m_p>\eta_{cr}$. The functions are plotted
at equal time steps with the higher amplitude curves occurring at later times.}
\label{n=1metric}
\end{figure}

\begin{figure}
\centerline{
\epsfbox{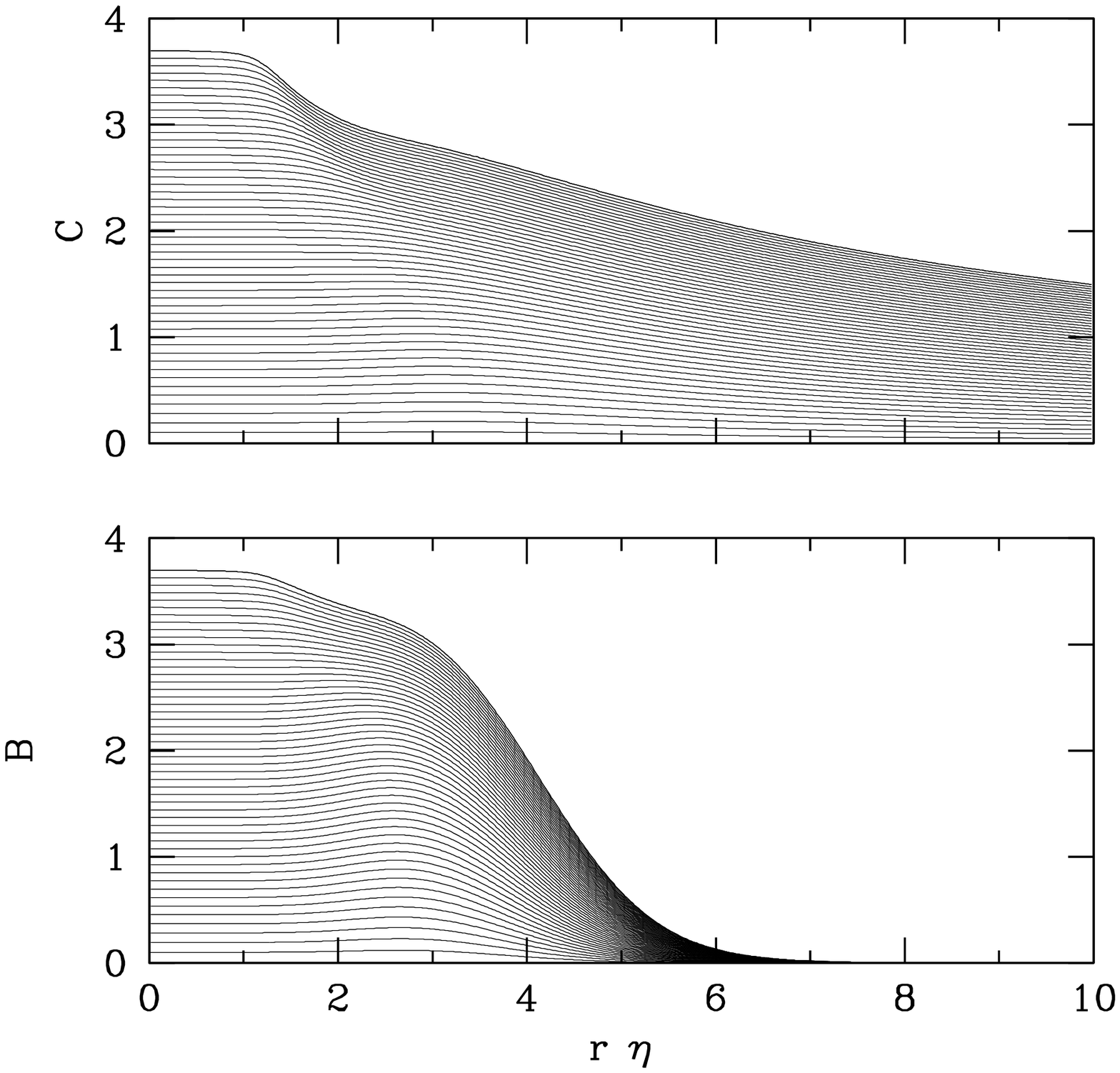}
}
\vspace{0.1in}
\caption{The metric fields $B(r)$ and $C(r)$ for an $n=5$ string with 
symmetry breaking scale $\eta=0.07m_p>\eta_{cr}$. The functions are plotted
at equal time steps with the higher amplitude curves occurring at later times.}
\label{n=5metric}
\end{figure}

As we mentioned in section II, our strategy was to evolve the initial 
configurations for a given value of the winding $n$, for various values of
the scale $\eta$. As an example, in Figure~(\ref{n=1metric}) we show the 
metric fields
$B(r,t)$ and $C(r,t)$ for an $n=1$ string with symmetry breaking scale
$\eta=0.2m_p>\eta_{cr}$. For comparison, in Figure~(\ref{n=5metric}) we show
the same fields for an $n=5$ string with $\eta=0.07>\eta_{cr}$.
These plots demonstrate that the metric fields grow exponentially in the
core but that the core size decreases exponentially. It is the competition
between these two effects that determines whether inflation occurs or not. 

In both
cases inflation is occurring in the core of the defect, although this is
not clear until we apply our criterion that the total volume of physical
space be increasing exponentially in the core. Note that, as an artifact of 
our initial conditions, there is an initial period of time during which the
system relaxes to its final state.

\begin{figure}
\centerline{
\epsfbox{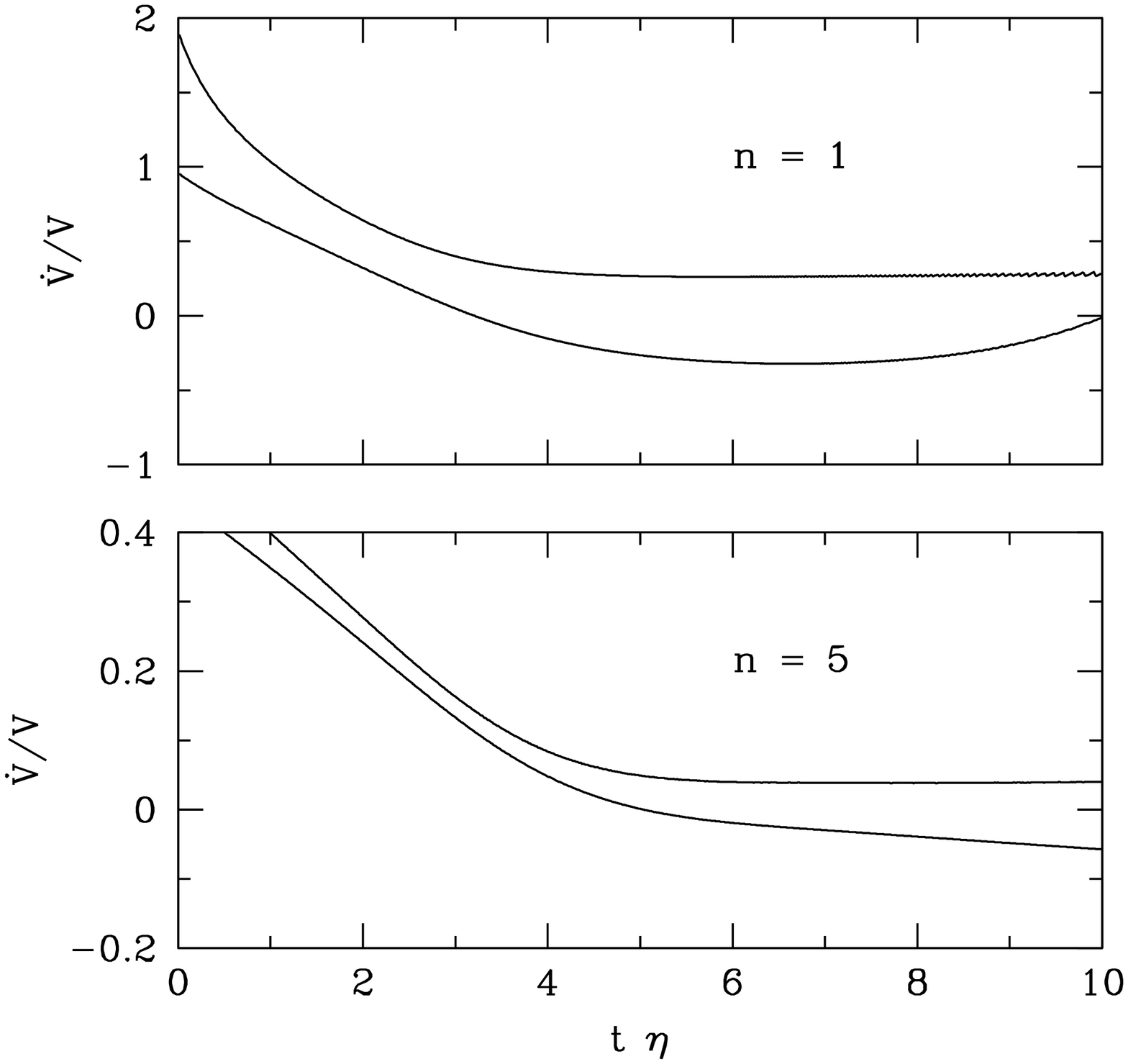}
}
\vspace{0.1in}
\caption{${\dot V}/V$ as a function of time $t$ for two values of $\eta$, one
for which the core inflates (upper curve in each figure), and the other for 
which it does not (lower curve in each figure), for two cases of the topological 
winding, $n=1$ and $n=5$}
\label{criterion}
\end{figure}

To illustrate how the criterion is applied, Figure~(\ref{criterion})
shows ${\dot V}/V$ as a function of time $t$ for two values of $\eta$, one
for which the core inflates, and the other for which it does not, for two
cases of the topological winding, $n=1$ and $n=5$. It is the
qualitative difference between these two classes of scales that allows us to 
home in on $\eta_{cr}\simeq 0.16$ for $n=1$ and $\eta_{cr}\simeq 0.06$ for
$n=5$. In the $n=1$ case, the lower curve appears to turn up at late times.
We believe this to be due to our rigid definition of the core radius, which 
does not take account of oscillations that appear in the matter fields. 
However, finite-size effects in the simulations limit our ability to test 
this.

\begin{figure}
\centerline{
\epsfbox{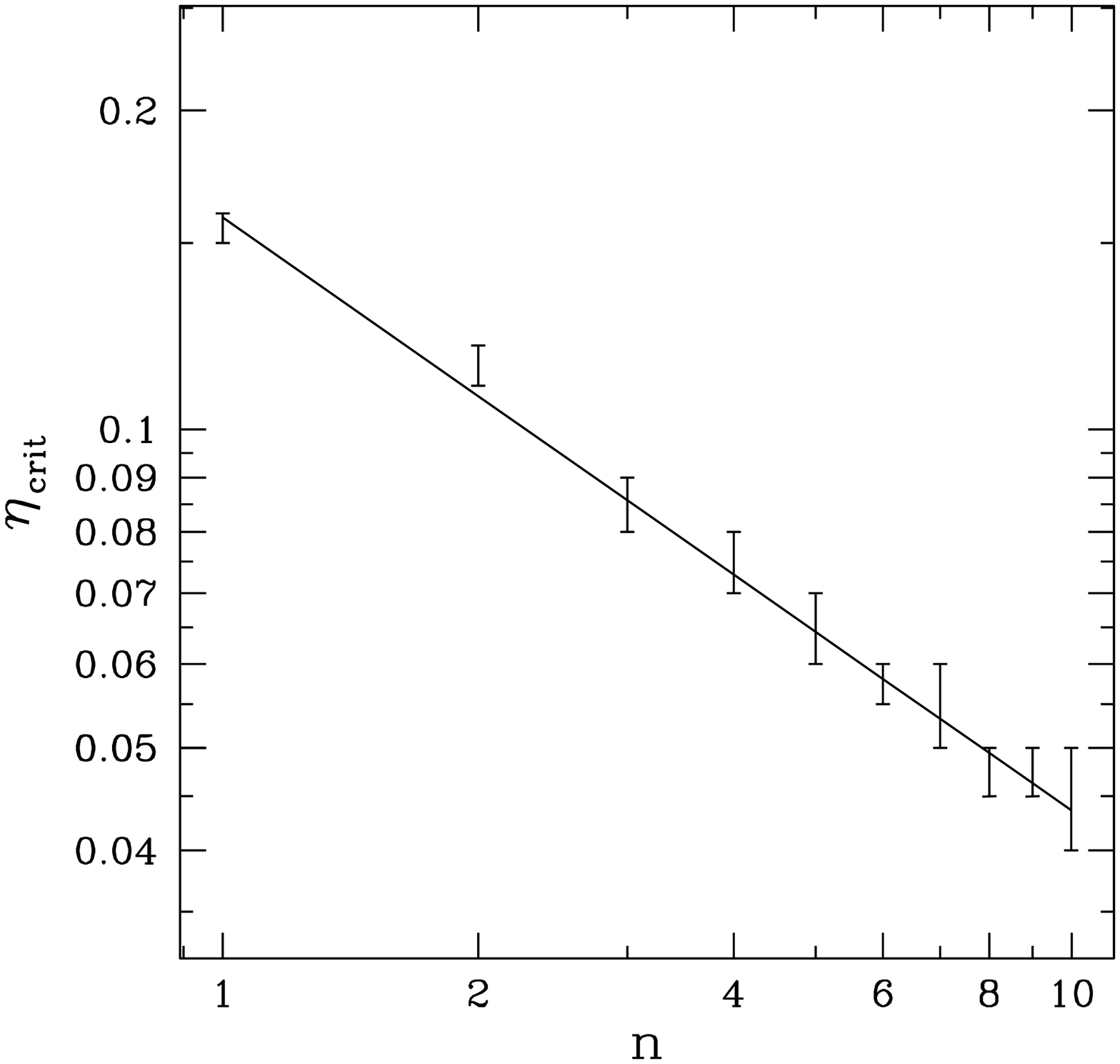}
}
\vspace{0.1in}
\caption{A log-log plot of $\eta_{cr}$, in units of $m_p$, versus the 
topological charge $n$. The ``error bars'' denote the ranges within which we
could numerically bracket $\eta_{cr}$. Their differing sizes reflect our
initial trial and error guesses for the bracketing.}
\label{etavsn}
\end{figure}

Finally, in Figure~(\ref{etavsn}) we show the relationship between
$\eta_{cr}$ and the topological charge $n$ on a log-log plot. The points are
best fit by a linear relationship

\be
\eta_{cr} \simeq \alpha n^p \ ,
\ee
where 

\be
\alpha = 0.16 \ , \ \ \ p = -0.56 \ .
\ee
This is in excellent agreement with the naive estimate $p=-0.5$ obtained by
analyzing the point at which the static asymptotic metric ceases to exist.

\section{Conclusions and Discussion}

We have analyzed the onset of topological inflation in the cores of
cosmic string solutions to the Einstein-Abelian-Higgs system in 
$(2+1)$ dimensions. For a soliton in a given sector of topological
charge $n$, inflation occurs in the defect core if the symmetry
breaking scale $\eta$ is greater than a critical value $\eta_{cr}(n)$.
The functional dependence of $\eta_{cr}$ on the winding $n$ was
determined numerically and was found to be monotonically decreasing
roughly, though not exactly, as $1/\sqrt{n}$. 
This result supports the intuition about defects with 
multiple winding gained from the asymptotic metric of static solutions
and from perturbative analyses of the core fields. If our results can
be extrapolated to very large $n$, and if strings with such high
winding form in phase transitions, then it is possible that
topological inflation could occur in GUT scale defects.

The present analysis is especially relevant in theories in which particles 
are viewed as solitons. In these theories one would expect that, at high
densities, the solitonic nature of particles would become important. Our
results then show that, provided the number of particles is large enough,
the tightly squeezed state of particles can start inflating. This may be
relevant for the gravitational collapse of stars since the number of 
particles in a star is of order $10^{57}$. In the context of the dual 
standard model, all particles correspond to magnetic monopoles and so
we would expect the present considerations to apply there also. However,
in this model, baryon number is not a conserved quantity
and it is possible that the star evaporates before inflation can set in, 
much like in the scenarios recently considered in \cite{upen,ms}.

A number of related investigations are suggested by our analysis.
First, is it possible that the collision of a monopole and an antimonopole
can result in an inflating region? Some years ago, 
Farhi, Guth and Guven \cite{fgg} considered the possibility of creating
a universe in particle collisions. Is their (negative) conclusion applicable
even in soliton collisions? Second, we have only considered strings at
critical coupling. For different choices of couplings,
the strings could attract or repel each other. How do our results depend
on the coupling constants? Does the instability of higher winding strings
to decay into those of lower winding, come into play and terminate topological
inflation after a certain number of e-folds?
We hope to return to some of these questions in future investigations.

\acknowledgements

We would like to thank Inyong Cho, Martin Lemoine, Glenn Starkman and
Alex Vilenkin for helpful discussions. A.D. and T.V. are thankful to the 
U.S. Department of Energy (D.O.E.) for support. The work of M.T. was supported 
by the D.O.E., the National Science foundation, and by funds provided by Case 
Western Reserve University.

\end{document}